\documentclass[letterpaper]{article} % DO NOT CHANGE THIS
\usepackage{aaai25}  % DO NOT CHANGE THIS
\usepackage{amssymb, latexsym, amsmath, mathtools, amsthm, mathrsfs} 
\usepackage{booktabs} 
\usepackage{times}  % DO NOT CHANGE THIS
\usepackage{helvet}  % DO NOT CHANGE THIS
\usepackage{courier}  % DO NOT CHANGE THIS
\usepackage[hyphens]{url}  % DO NOT CHANGE THIS
\usepackage{graphicx} % DO NOT CHANGE THIS
\usepackage{natbib}  % DO NOT CHANGE THIS AND DO NOT ADD ANY OPTIONS TO IT
\usepackage{caption} % DO NOT CHANGE THIS AND DO NOT ADD ANY OPTIONS TO IT

\usepackage{algorithm}
\usepackage{algorithmic}

\usepackage{newfloat}
\usepackage{listings}
\DeclareCaptionStyle{ruled}{labelfont=normalfont,labelsep=colon,strut=off} % DO NOT CHANGE THIS
\floatstyle{ruled}
\newfloat{listing}{tb}{lst}{}
\floatname{listing}{Listing}
\title{MMVA: Multimodal Matching Based on Valence and Arousal across Images, Music, and Musical Captions }
\author {
    Suhwan Choi\textsuperscript{\rm 1,2}
    Kyu Won Kim,
    Myungjoo Kang\textsuperscript{\rm 1}
}
\affiliations {
    \textsuperscript{\rm 1}Seoul National University, South Korea\\
    \textsuperscript{\rm 2}CRABs\\
    schoi@crabs.ai, qkim828@gmail.com, mkang@snu.ac.kr
    }
\begin{document}

\maketitle

\begin{abstract}
We introduce Multimodal Matching based on Valence and Arousal (MMVA), a tri-modal encoder framework designed to capture emotional content across images, music, and musical captions. To support this framework, we expand the Image-Music-Emotion-Matching-Net (IMEMNet) dataset, creating IMEMNet-C which includes 24,756 images and 25,944 music clips with corresponding musical captions. We employ multimodal matching scores based on the continuous valence (emotional positivity) and arousal (emotional intensity) values. This continuous matching score allows for random sampling of image-music pairs during training by computing similarity scores from the valence-arousal values across different modalities. Consequently, the proposed approach achieves state-of-the-art performance in valence-arousal prediction tasks. Furthermore, the framework demonstrates its efficacy in various zeroshot tasks, highlighting the potential of valence and arousal predictions in downstream applications.

\end{abstract}

% Uncomment the following to link to your code, datasets, an extended version or similar.

\begin{links}
    %\link{Code & Dataset}{https://aaai.org/example/code}
    %\link{Datasets}{https://aaai.org/example/datasets}
    %\link{Extended version}{https://aaai.org/example/extended-version}
\end{links}

\section{Introduction}
Integrating multiple modalities in machine learning, particularly image, audio, and text, presents a compelling yet complex challenge. This complexity stems from the inherent diversity and intricate relationships of these three modalities. AudioCLIP \cite{Audioclip9747631} has established foundational work in this domain, capturing cross-modal relationships between non-musical audio, texts, and images. However, prior research did not specifically address music's interplay with associated visual and textual components.

Building upon this groundwork, we address a novel problem by focusing on the multimodal relationships between images, music, and text. To the best of our knowledge, this work is the first in multimodal learning to specifically examine the interplay among these three modalities, particularly in the context of musical captions. By shifting the focus to this underexplored combination of modalities, we aim to open new research avenues that explore how these three modalities can interact to generate meaningful insights and applications. 
\begin{figure}[t]
\centering
\includegraphics[width=1.\linewidth]{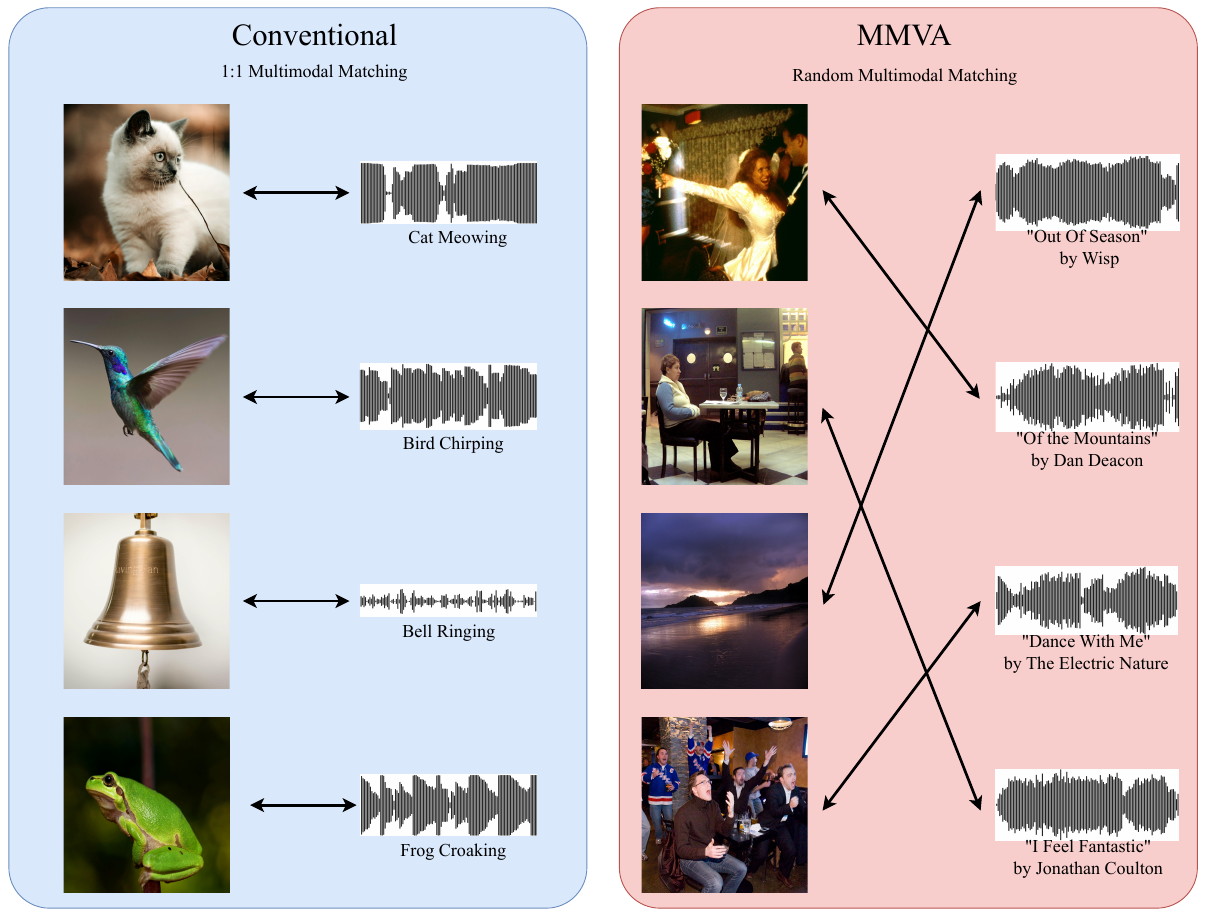}
%\vspace{-0.7cm}
\caption{Comparison of multimodal matching.}
%\vspace{-0.5cm}
\label{fig:multimodal_matching}
\end{figure}
To further this exploration, we expand the existing Image-Music-Emotion-Matching-Net (IMEMNet) \cite{CDCML} by incorporating musical captions aligned with their audio counterparts. This expansion not only enriches the dataset but also challenges the conventional approach in multimodal research, which typically relies on one-to-one matching pairs between different modalities. For example, in AudioCLIP, an image of a cat is precisely matched with the audio of a cat meowing and the text ``cat.'' In contrast, the IMEMNet dataset deviates from this conventional paradigm. Instead of direct correlations between modalities, it features image-music matching based on continuous matching scores derived from valence (emotional positivity) and arousal (emotional intensity). 

Leveraging IMEMNet, this research navigates a more nuanced landscape where the degree of multimodal correlation is variable and quantifiable only through a matching score. While previous work \cite{CDCML} using the IMEMNet dataset employed predefined multimodal pairs, the current study implements random sampling of multimodal pairs during training. This approach yields superior performance on the valence-arousal (VA) prediction task compared to existing works, while also demonstrating the applicability of valence and arousal to various zero-shot tasks previously unexplored. Extensive experiments validate the effectiveness of the proposed method. This research thus not only introduces a novel set of modalities into the multimodal learning paradigm but also broadens the scope for potential applications of these combined modalities.

\section{Related Work}
\paragraph{Multimodal learning with tower encoders.}
In multimodal learning, a tower encoder refers to a separate neural network architecture that processes data of a particular modality, as exemplified by the CLIP image and text encoder of CLIP \cite{CLIP}. Tower-based multimodal learning offers several advantages. First, each tower can be optimized for its specific data type, leveraging architectures best suited to the unique characteristics of the particular data modality. Second, it allows for the independent scaling and improvement of each tower without necessitating a redesign of the entire model. Updates or changes can be made to one modality without impacting the others.

Previous research has demonstrated that the audio modality can be successfully integrated with other modalities in multimodal learning using tower encoders. Conventional audio-text multimodal research \cite{wu2023large, CLAP2022, Mei2022metric} requires precise one-to-one audio-text matching in the speech data. Likewise, training a three-tower framework \cite{Audioclip9747631} capable of processing audio, image, and text relies on precise multimodal matching, as shown in Figure \ref{fig:multimodal_matching}. 

\paragraph{Multimodal learning with music.}
Image-music multimodal learning remains largely unexplored, in contrast to music-text multimodal learning. For music-text tasks, MuLan \cite{huang2022mulan} trains a text encoder and a music encoder in a two-tower framework for music tagging and music-text retrieval. For music-image tasks, CDCML \cite{CDCML} constructed IMEMNet, a dataset of image-music pairs with VA matching scores, and proposed a multimodal training framework for emotion-based image-music matching. Employing a one-to-one image-music 
matching dataset, \cite{nakatsuka2023content} jointly trains image and music tower encoders to retrieve music from album covers and vice versa. 
\section{Dataset}
The biggest challenge to multimodal learning involving image, music, and musical caption is data scarcity. To address this, IMEMNET-C extends the original IMEMNet dataset \cite{CDCML} by incorporating musical captions corresponding to the music data in IMEMNet. 
\subsection{IMEMNet}
IMEMNet integrates a music dataset with a diverse range of image datasets, each annotated with distinct emotional valence-arousal (VA) labels, to establish a comprehensive dataset for multimodal studies.

\paragraph{Image datasets.} IMEMNet incorporates three distinct real-world image datasets: IAPS \cite{IAPS}, NAPS \cite{NAPS}, and EMOTIC \cite{EMOTIC}. The International Affective Picture System (IAPS) comprises 1,182 images, each annotated using a 9-point Valence-Arousal-Dominance (VAD) scale. The Nencki Affective Picture System (NAPS) includes 1,356 images evaluated on a 9-point bipolar semantic sliding scale, focusing on VA dimensions and approach-avoidance measures. The EMOTions In Context (EMOTIC) features 23,082 images of people, with annotations on a continuous 10-point VAD scale.

\paragraph{Music dataset.} IMEMNet utilizes the DEAM (Database for Emotion Analysis using Music) dataset, which consists of 1,802 music tracks, each annotated with valence and arousal values ranging from -1 to +1. These annotations are provided on a per-second basis and for the entire track. To ensure annotation stability, the annotation for each track starts at the 15th second. The tracks, predominantly 45 seconds in length and recorded at a frequency of 44100Hz, vary in duration, with some extending over 600 seconds. By dividing these tracks into two-second segments, the dataset comprises a total of 23,944 clips. Unlike \citet{CDCML}, we extend the maximum clip length to 10 seconds by concatenating up to five consecutive clips with each two-second clip, allowing our framework to handle longer music data.

\paragraph{Combining the datasets.} To address the different scales used for labeling the image and music data, VA values are normalized to a [0,1] range. In total, IMEMNet comprises 25,620 images and 23,944 music clips, resulting in 144,435 image-music pairs, which are divided into 80\% for training, 5\% for validation, and 15\% for testing.

\paragraph{Obtaining matching scores.} To align images and music clips, image-music matching scores are calculated based on the Euclidean distance between their two-dimensional VA vectors. The similarity between an image $I_i$ and a music clip $M_j$ is quantified by the following:
\begin{equation}\label{eq:matching_score}
S(I_i,M_j) = \exp (-\frac{d(y^{I_i}, y^{M_j})}{\sigma_a^b}),
\end{equation}
where $a$ and $b$ represent the total number of images and music clips respectively, such that $i \in \{1, ... , a\}$ and $j \in \{1,...,b\}$. Here, $d(y^{I_i}, y^{M_j})$ denotes the Euclidean distance between the VA labels of image $I_i$ and music clip $M_j$, and $\sigma_a^b$ is the average Euclidean distance across all image and music clip pairs. This similarity measure serves as the emotional matching label for each corresponding image and music clip pair. 

It's important to highlight that the potential number of all matching pairs in this setup is $a\times b$. Given the scale of the image and music clip datasets, this could result in hundreds of millions of possible pairs. To manage this vast number and prevent an overwhelming expansion of the dataset, \citet{CDCML} chose 50 images for each music clip: 30 are randomly picked from the image dataset, and the remaining 20 are split equally between those with the highest and lowest matching scores. To further refine the dataset, 10\% of these pairs are randomly sampled to form the final IMEMNet dataset. \citet{CDCML} also ensured no overlap between the training, validation, and test sets. 

\subsection{IMEMNet-C}\label{dataset:caption} 
The IMEMNet dataset only contains image and music pairs, lacking musical captions for the music data. To address this gap, we introduce a new dataset, IMEMNet-C, which contains musical captions corresponding to the music data. 
Music captions are generated by the music-to-text large language model, LP-MusicCaps \cite{musiccaps}. 

Evaluations of the generated captions revealed the prevalence of redundant phrases, such as ``The low quality recording.'' To ensure high-quality captions for the music-image pairs, redundant phrases were removed in two steps. First, a rule-based refinement was applied to part of the audio captions. About half of the captions with redundant phrases contained the exact phrase ``low quality recording features.'' This redundant phrase was rephrased as ``recording features.'' Then, Llama-3.1-8B \cite{LLAMA3} refined musical captions by removing and rephrasing the redundant expressions. We use the prompt ''Refine or remove an audio quality-related phrase to eliminate any mention of audio quality'' with 4-shot examples.

\section{The Proposed Method: MMVA}

\paragraph{Asymetric multimodal matching.} Multimodal training with IMEMNet-C presents a novel scenario in the field of multimodal learning. In this setup, multimodal matching exhibits asymmetry: some modality pairs (music-text) are one-to-one matched, while others (image-text, image-music) are not. For instance, each music clip has its corresponding musical caption, just as in conventional multimodal training scenarios. Accordingly, music clips and their corresponding captions share the same VA values. However, the multimodal pairs involving images, such as image-music and image-caption, lack exact correspondence. In these cases, the degree of matching is represented by continuous emotional matching scores ranging from 0 to 1, providing a nuanced approach to multimodal alignment.

\paragraph{Random multimodal matching.} Figure \ref{fig:multimodal_matching} illustrates the difference between conventional multimodal matching and random multimodal matching. Conventional multimodal matching relies on one-to-one matching between multimodal data, whereas our approach does not. The continuous nature of valence and arousal allows any multimodal pair to have a similarity matching score using Equation \ref{eq:matching_score}. As long as valence and arousal values are available, this dramatically reduces the burden of obtaining multimodal data.
\begin{figure}[t]
\centering
\includegraphics[width=1.\linewidth]{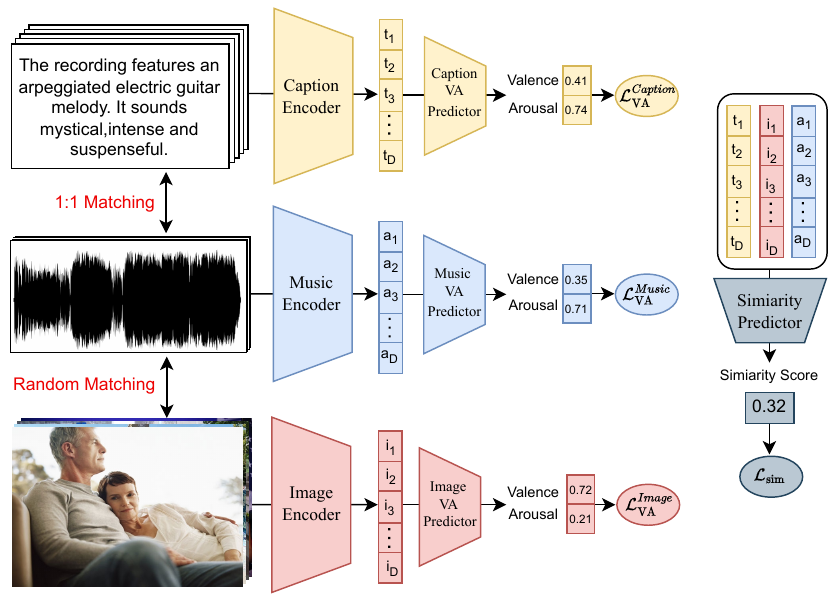}
\caption{Overview of MMVA.}
\label{fig:overview}
\end{figure}
\subsection{Random Matching Based on Valence and Arousal}\label{method_loss}
To handle the asymmetric multimodal training scenario, we propose \textbf{MMVA}, a framework that leverages random multimodal matching to facilitate learning across images, music, and musical captions. During training, the framework uses Valence-Arousal (VA) predictions to preserve the emotional integrity of each modality while optimizing similarity scores to align modalities cohesively in the absence of precise matches. Figure \ref{fig:overview} illustrates the overview of MMVA. 

Conventional multimodal learning approaches, such as AudioCLIP, rely on contrastive loss to align one-to-one paired data modalities and optimize semantic similarity by comparing high-dimensional embeddings. In contrast, MMVA aligns three modalities through emotional coherence in asymmetric scenarios where exact pairings are unavailable. Instead of comparing high-dimensional embeddings, MMVA optimizes low-dimensional VA (Valence-Arousal) vectors against ground truth VA values. To achieve this, we employ two types of loss functions based on Mean Squared Error (MSE): single-modality VA prediction loss ($\mathcal{L}_{\text{VA}}$) and multimodal similarity prediction loss ($\mathcal{L}_\text{sim}$). 

\paragraph{VA prediction loss.} For each embedding vector $\boldsymbol{z}^k$ corresponding to modality $k$, we obtain a two-dimensional VA vector $\boldsymbol{e}^k =[valence,arousal]^\top$ such that $\boldsymbol{e}^k = p_k(\boldsymbol{z}^k)$, where $p_k$ is the VA predictor for the modality $k$. Then, for a batch of VA predictions $E^k$ and a corresponding batch of ground-truth VA vectors $Y_{\text{VA}}^k$ for modality $k$, we compute single-modality VA prediction loss: 
\begin{equation}\label{eq:va_loss}
    \mathcal{L}_{\text{VA}}^k= \frac{1}{N}\sum_{n=1}^N||\boldsymbol{E}^k_n - \boldsymbol{Y}_n^k||^2_2.
\end{equation}
$\mathcal{L}_{\text{VA}}^k$ quantifies the accuracy of the VA predictions for each modality, providing a measure of how well the model's predictions align with the actual emotional valence and arousal levels represented in the data. By minimizing this loss during training, the model learns to generate more emotionally accurate and representative embedding $\boldsymbol{z}$ for each modality.
\begin{table*}[t]
    \centering
    %\vskip -0.05in
    %\vspace{-10pt}
    \resizebox{1\textwidth}{!}{
    \begin{tabular}{lcc|cccc|cccc|cccc}\\\toprule  
     &\multicolumn{2}{c}{\textbf{Simiarity}}&\multicolumn{4}{c}{\textbf{Image Emotion}} & \multicolumn{4}{c}{\textbf{Music Emotion}} & \multicolumn{4}{c}{\textbf{Text Emotion}}\\
     &\multicolumn{1}{c}{\textit{MSE}}&\multicolumn{1}{c}{\textit{MAE}}&\multicolumn{1}{c}{\textit{V MSE}}&\multicolumn{1}{c}{\textit{V MAE}}&\multicolumn{1}{c}{\textit{A MSE}}&\multicolumn{1}{c}{\textit{A MAE}} &\multicolumn{1}{c}{\textit{V MSE}}&\multicolumn{1}{c}{\textit{V MAE}}&\multicolumn{1}{c}{\textit{A MSE}}&\multicolumn{1}{c}{\textit{A MAE}} &\multicolumn{1}{c}{\textit{V MSE}}&\multicolumn{1}{c}{\textit{V MAE}}&\multicolumn{1}{c}{\textit{A MSE}}&\multicolumn{1}{c}{\textit{A MAE}}  \\\midrule
     SP-Net & 0.135 & 0.301 & 0.048 & 0.165 & 0.054 & 0.186 & 0.026 & 0.120 & 0.020 & 0.114 & \\\midrule\addlinespace[0.1cm]
    $L^3$-Net & 0.095 & 0.232 & 0.058&0.183 &0.085 &0.232 &0.034 &0.143 & 0.028 &0.136 &  \\\midrule\addlinespace[0.1cm]
    ACP-Net & 0.086 &0.222 & 0.062&0.195 &0.091 &0.241 &0.027 &0.130 &0.022 & 0.131&  \\\midrule\addlinespace[0.1cm]
    CDCML & \underline{0.067} &\underline{0.210} & \underline{0.044} &\underline{0.157} & \underline{0.050}&\textbf{0.175} &\underline{0.024} &\underline{0.118} & \underline{0.015} &\underline{0.099} &  \\\midrule\addlinespace[0.1cm]
    \textbf{MMVA}  &\textbf{0.033} &\textbf{0.147} &\textbf{0.0231} &\textbf{0.110} & \textbf{0.049} & \underline{0.181} & \textbf{0.0003} &\textbf{0.008} & \textbf{0.0002} &\textbf{0.007} &0.019 &0.094 &0.015& 0.082 \\
    \bottomrule
    \end{tabular}}
    \caption{Performance of VA prediction. The best results are emphasized in bold, while the second-best ones are marked with an underline.}\label{table:va_results}
    %\vskip -0.1in %0.5
\end{table*}
\paragraph{Similarity matching loss.} Each triplet, comprising an image, a music clip, and a musical caption, is assigned a single emotional matching score. Since music clips and their corresponding captions share identical VA values, an emotional matching score of a triplet directly corresponds to the similarity score between the image and the music clip. 

For each triple of embedding vectors, we generate a unified multimodal embedding vector by concatenating the individual embeddings: $\boldsymbol{m}=\text{concat}(\boldsymbol{z}^{Img},\boldsymbol{z}^{Mus}, \boldsymbol{z}^{Cap})$. A similarity predictor $p_{\text{sim}}$ then maps this multimodal embedding vector $\boldsymbol{m}$ to an emotional matching score within the range [0,1]. Then, given a batch of predicted scores $S$ and a corresponding batch of ground-truth scores $Y_{\text{sim}}$, similarity matching loss is as follows: 
\begin{equation}\label{eq:sim_loss}
    \mathcal{L}_{\text{sim}}= \frac{1}{N}\sum_{n=1}^N(S_n - Y_n^{sim})^2.
\end{equation}
$\mathcal{L}_{\text{sim}}$ evaluates the emotional congruence within a triplet, assessing how well the emotional content of the image aligns with the combined emotional essence of the music and its caption. %Moreover, by minimizing this loss, the model becomes more adept at understanding and quantifying the complex emotional relationships between images, music, and captions in the multimodal setting.

\paragraph{Training objective.} During training, we optimize the following objective, assigning an equal weight of 1 to each loss:
\begin{equation}\label{eq:total_loss}
    \mathcal{L}_{\text{MMVA}}= \mathcal{L}_{\text{VA}}^{Img}+\mathcal{L}_{\text{VA}}^{Mus}+\mathcal{L}_{\text{VA}}^{Cap}+\mathcal{L}_{\text{sim}}
\end{equation}
By minimizing $\mathcal{L}_{\text{MMVA}}$, the models illustrated in Figure \ref{fig:overview} learn to quantify the complex emotional relationships between images, music, and captions.
\section{Experiments}
\subsection{Training Details}
The AdamW \cite{adamw} optimizer is utilized with a learning rate of $0.0003$, a weight decay of $0.01$, $\beta_1$ of $0.9$, and $\beta_2$ of $0.999$. The training is conducted for 1000 epochs, using a batch size of 128 and a cosine annealing schedule \cite{cosanneal}. 

\subsection{Architectures}\label{architecture}
As depicted in Figure \ref{fig:overview}, MMVA utilizes 7 separate modules. These include a modality-specific encoder $f_k$ and VA predictor $p_k$ for $k \in \{Image, \ Music, \ Caption \}$, along with a similarity predictor encoder $p_\text{sim}$.
\paragraph{Modality-specific encoder.} For the image encoder $f_{img}$, MMVA employs the CLIP image encoder \cite{CLIP}, which is based on ViT-B/32 \cite{vit}. $\boldsymbol{z}^{Img}$ corresponds to the $512$-dimensional class token vector. For both music and captions, MMVA uses BERT-based models \cite{BERT}: MERT-95M \cite{MERT} for $f_{mus}$ and RoBERTa-base \cite{roberta} for $f_{cap}$. For each of these BERT-based models, the proposed method extracts class tokens vectors $\boldsymbol{h}_l\in \mathbb{R}^{768}$ from the last $13$ transformer blocks \cite{transformer} and gathers the vectors to form modality features $\boldsymbol{h} \in \mathbb{R}^{13 \times 768}$. Subsequently, a 1D convolution is applied to derive a $768$-dimensional modality vector, which is then linearly projected to obtain the modality-specific embedding $\boldsymbol{z}^k \in \mathbb{R}^{512}$.  

\paragraph{Emotion predictors.}\label{emotion_predictor_arch}
Mapping $\boldsymbol{z}^k$ to two-dimensional VA prediction vectors, VA predictors $p_\text{VA}^k$ share the same architecture across $k$: \textbf{linear512-relu-layernorm-dropout0.5-linear512-relu-layernorm512-dropout0.5-linear2-sigmoid}. The similarity predictor $p_\text{sim}$ is identical to $p_\text{VA}^k$ except the first and last layers because $p_\text{VA}^k$ takes $1536$-dimensional multimodal embedding vectors $m$ as inputs and outputs a one-dimensional value.

\subsection{VA Prediction}
MMVA integrates emotion prediction into its training process by utilizing the VA predictors, $p_\text{VA}^k$, and similarity predictor, $p_\text{SIM}$. These modules are designed to capture emotional content by learning to predict valence-arousal values. VA prediction performance illustrates how well modality-specific emotional information is preserved during multimodal training. On the other hand, similarity prediction performance reflects the degree of multimodal information present in the embeddings. Together, these metrics provide a comprehensive assessment of the overall efficacy of the multimodal training process.

\paragraph{Setups.}
We compare MMVA against the baseline results obtained by \citet{CDCML}: SP-Net, $L^3$-Net \cite{L3-NET}, ACP-Net \cite{ACP-Net}, and CDCML \cite{CDCML}. All the baselines are tailored for emotion-based image-music correspondence, employing the same emotion predictor architectures as used in our work. Evaluation employs Mean Squared Error (MSE) and Mean Absolute Error (MAE) as metrics to quantify the similarity between the predicted and ground-truth emotion values.

\paragraph{Results.}
Table \ref{table:va_results} presents the emotion prediction performance across all modalities. Because the baselines were trained on image-music modalities only, they do not provide VA performance metrics for captions. For image and music modalities, MMVA outperforms the best baseline CDCML across all metrics except MAE of image arousal.

\begin{figure}[t]
\centering
\includegraphics[width=\linewidth]{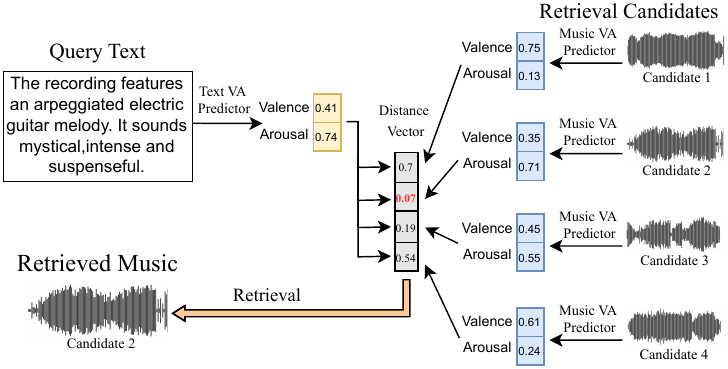}
\caption{Overview of text-to-music cross-modal retrieval.}
\label{fig:retrieval_overview}
\end{figure}

\subsection{Text-to-Music Retrieval}
\begin{table}[h]
%\vspace{-13pt}%-12
  \centering
  \resizebox{0.47\textwidth}{!}{
    \begin{tabular}{@{}lrrrr@{}}
      \toprule
      &\multicolumn{4}{c}{\textbf{1000 Music-Captions}} \\
      \cmidrule(r){1-5}
      & R@1   &R@5 &R@10   &mAP10 \\\addlinespace[0.1cm]   
      \midrule
      \addlinespace[0.1cm]
      Class-bin \cite{classbin}  &4.0   &13.8   & 20.1    &8.3 \\    \addlinespace[0.1cm]  
      \midrule 
      Triplet-Glove \cite{triplet-glove} &2.8   &11.2   & 18.6    &6.6 \\    \addlinespace[0.1cm]   
      \midrule
      Triplet-BERT  \cite{triplet-bert}&\underline{6.7}   &23.6   & 36.6    &14.1 \\    \addlinespace[0.1cm]   
      \midrule
      Contrastive-BERT \cite{contrastive-bert} &\textbf{10.2}   &\textbf{29.8}   & \textbf{42.8}    &\textbf{18.7} \\    \addlinespace[0.1cm]   
      \midrule
      \textbf{MMVA} (IMEMNet-C) &  \underline{6.7}  &18.6 &  24.4   &11.66  \\
      \bottomrule
    \end{tabular}
  }
\caption{Text-Music Retrieval}\label{retrieval_result}
%\vspace{-15pt}
\end{table}

Leveraging the VA-based similarity score (Equation \ref{eq:matching_score}), MMVA can achieve cross-modal retrieval for various modality pairs. Figure \ref{fig:retrieval_overview} illustrates the text-to-music retrieval based on valence and arousal. Given a query text, the music clip with the highest similarity score is retrieved. Table \ref{retrieval_result} demonstrates that MMVA achieves competitive retrieval performance. It is important to note that the baseline methods retrieve music clips from the Million Song Dataset \cite{millionsong}, which provides only pre-extracted audio features. In contrast, the MMVA music encoder operates directly on raw audio, making direct comparisons challenging due to differences in input representation and data.

\subsection{Image-to-Text Retrieval for Music Generation.}
Leveraging the VA-based retrieval process, MMVA can generate text prompts for text-to-music generative models.
\begin{figure}[t]
\centering
\includegraphics[width=\linewidth]{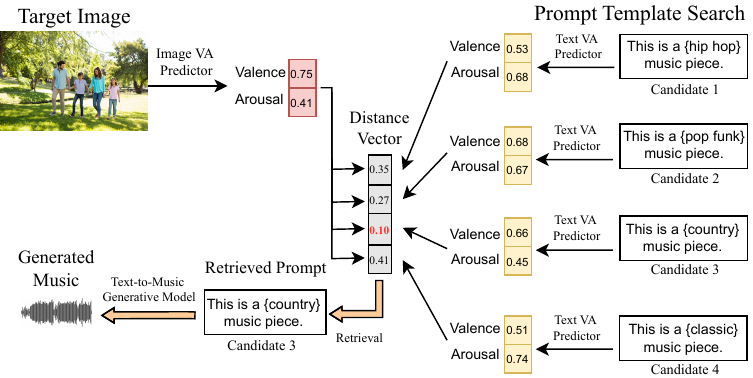}
\caption{Overview of Image-to-Music generation.}
\label{fig:prompt_generation}
\end{figure}

\begin{figure}[t]
\centering
\includegraphics[width=\linewidth]{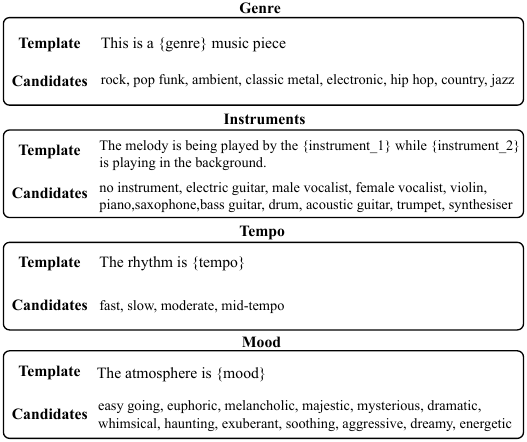}
\caption{Musical prompt templates.}\label{prompt_template}
\end{figure}

\begin{figure}[t]
\centering
\includegraphics[width=\linewidth]{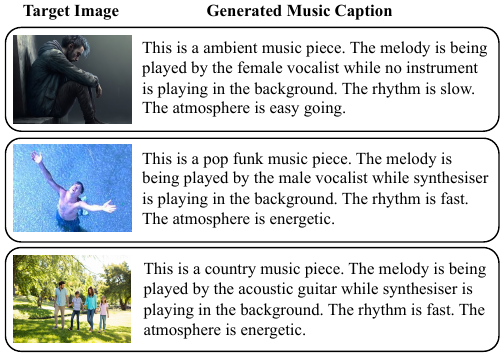}
\caption{Captions generated from images by MMVA.}\label{generated_captions}
\end{figure}
\paragraph{Music generation.} Figure \ref{fig:prompt_generation} illustrates the image-to-text generation processes. First, a VA vector for the target image is obtained. Then, text prompts are exhaustively generated by inserting candidates into each text template illustrated in Figure \ref{prompt_template}. As a result, VA values for the generated text prompts are obtained, allowing for the calculation of the VA distance between the target image and the text prompts. By selecting the text prompt with the smallest distance, the best text prompt for the image is identified and fed into the text-to-music generative model, e.g. MusicGen \cite{musicgen}. Generated captions are illustrated in Figure \ref{generated_captions}. 
\begin{table*}[h]
    \centering
    %\vskip -0.05in
    %\vspace{-10pt}
    \resizebox{1\textwidth}{!}{
    \begin{tabular}{lcc|cccc|cccc|cccc}\\\toprule  
     &\multicolumn{2}{c}{\textbf{Simiarity}}&\multicolumn{4}{c}{\textbf{Image Emotion}} & \multicolumn{4}{c}{\textbf{Music Emotion}} & \multicolumn{4}{c}{\textbf{Text Emotion}}\\
     &\multicolumn{1}{c}{\textit{MSE}}&\multicolumn{1}{c}{\textit{MAE}}&\multicolumn{1}{c}{\textit{V MSE}}&\multicolumn{1}{c}{\textit{V MAE}}&\multicolumn{1}{c}{\textit{A MSE}}&\multicolumn{1}{c}{\textit{A MAE}} &\multicolumn{1}{c}{\textit{V MSE}}&\multicolumn{1}{c}{\textit{V MAE}}&\multicolumn{1}{c}{\textit{A MSE}}&\multicolumn{1}{c}{\textit{A MAE}} &\multicolumn{1}{c}{\textit{V MSE}}&\multicolumn{1}{c}{\textit{V MAE}}&\multicolumn{1}{c}{\textit{A MSE}}&\multicolumn{1}{c}{\textit{A MAE}}  \\\midrule
    \textit{No Random Matching}  &0.085 &0.247 &0.051 &0.174 & 0.065 &0.208 & 0.049 &0.168 & 0.014 &0.092 &0.053 &0.179 &0.021& 0.118 \\\midrule\addlinespace[0.1cm]
    \textit{No Similarity Predictor}  &- &- &0.0232 &0.114 &0.053 & 0.185 & 0.0004 &0.010 & 0.0003  &0.010 &0.020 &0.102 &0.017& 0.091 \\\midrule\addlinespace[0.1cm]
    \textbf{MMVA}  &\textbf{0.033} &\textbf{0.147} &\textbf{0.0231} &\textbf{0.110} & \textbf{0.049} & \textbf{0.181} & \textbf{0.0003} &\textbf{0.008} & \textbf{0.0002} &\textbf{0.007} &\textbf{0.019} &\textbf{0.094} &\textbf{0.015}& \textbf{0.082} \\
    \bottomrule
    \end{tabular}}
    \caption{Ablation Performance of VA prediction. }\label{table:va_ablation}
    %\vskip -0.1in %0.5
\end{table*}

\subsection{Video Summarization}
Arousal represents the intensity of an emotion, with higher arousal levels indicating more intense emotional responses. Based on the premise that highlights in a video generally correspond to moments of high arousal, we can utilize arousal values to summarize videos. In the absence of a dedicated video summarization benchmark for music, we employ the image VA predictor, $p_{img}$, to identify significant frames within a video and benchmark its performance for image-based video summarization.

\paragraph{Setups.} To evaluate the effectiveness of arousal-based video summarization, we use the TVSum\cite{TVSUM} dataset, with frame-level importance scores assigned every two seconds. These scores range from 1 
(not important) to 5 (very important). MMVA's zeroshot approach is compared against well-established unsupervised video summarization methods: DR-DSN\cite{DR-DSN}, CSNet \cite{CSNet}, AC-SUM-GAN \cite{AC-SUM-GAN}, and CA-SUM \cite{CA-SUM}. For generating video summaries, MMVA follows the same procedure as these baselines, solving a knapsack problem to select frames based on the frame-level importance scores, with each frame representing a short clip in a video. The duration of the summarized videos remained within 15\% of the original.

\begin{table}[h]
\begin{center}
\begin{tabular}{lrc}
\toprule
 & F-score\\\midrule
 Random summary& 54.4 \\
\midrule
DR-DSN & 57.6\\
\midrule
CSNet & 58.8 \\
\midrule
AC-SUM-GAN & 60.6 \\
\midrule
CA-SUM & 61.4 \\
\midrule
\textbf{MMVA} & 58.2 \\
\bottomrule
\end{tabular}
\caption{Video summarization performance.}\label{table:video_sum}
\end{center}
%\vspace{-30pt}
\end{table} 

\paragraph{Evaluation.} 
F-score is used to evaluate the similarity between predicted summary $\hat{Q}$ and human-defined summary $Q$, given by:\begin{equation}
    \text{FS} = \frac{2(P\times R)}{P + R}, P = \frac{||\hat{Q} \cap Q||}{||\hat{Q}||}, R = \frac{||\hat{Q}\cap Q||}{||Q||} 
\end{equation} 
such that $P$ represents precision, and $R$ represents recall. Here, $||\cdot ||$ denotes duration and $\cap$ signifies temporal overlap. Unlike the baseline models trained on 90\% of the dataset and tested on the remaining 10\%, MMVA is evaluated on the entire dataset. Generating video summaries with MMVA is zeroshot, eliminating the need for a train/test split.

\paragraph{Results.}
Table \ref{table:video_sum} presents the video summarization performance. MMVA outperforms random summary and DR-DSN, showcasing promising performance, especially given that MMVA operates in a zero-shot setting.

\paragraph{Video summarization based on music.} Although the experiments for Table \ref{table:video_sum} were conducted using arousal values extracted from images, it is also feasible to summarize videos using arousal values derived from music. By segmenting a video into fixed-length clips, we can extract the arousal value for each clip and apply a knapsack problem formulation to select frames that exhibit high arousal, summarizing the video based on musical emotional intensity.

\subsection{Ablation Experiment}
\paragraph{Setup.} Two ablation setups are used for comparison. \textit{No Random Matching} is trained using the pre-defined multimodal pairs originally used in IMEMNet rather than the random multimodal matching. \textit{No Similarity Predictor} relies solely on modality-specific VA predictors, discarding the similarity predictor shown in Figure \ref{fig:overview}. Other factors, such as training hyperparameters, remain constant.
\paragraph{Results.}
\begin{table}[h]
%\vspace{-12pt}%-12
  \centering
  \resizebox{0.47\textwidth}{!}{
    \begin{tabular}{@{}lrrrrc@{}}
      \toprule
      &\multicolumn{4}{c}{\textbf{Text-Music Retrival}}&\multicolumn{1}{c}{\textbf{Video Sum.}} \\
      \midrule
      & R@1   &R@5 &R@10   &mAP10 & F-score \\\addlinespace[0.1cm]   
      \midrule
      \textit{No Random Matching}  &0.4   &1.4   & 2.9    &1.00 & 56.6\\    \addlinespace[0.1cm]   
      \midrule
      \textit{No Similarity Predictor} &5.8   &15.6   & 19.9    &10.00 & 55.9\\    \addlinespace[0.1cm]   
      \midrule
      \textbf{MMVA}  &  \textbf{6.7}  &\textbf{18.6} &  \textbf{24.4}   &\textbf{11.66} & \textbf{58.2} \\
      \bottomrule
    \end{tabular}
  }
  \caption{Ablation results of zeroshot performance.}\label{zeroshot_ablation}
%\vspace{-15pt}
\end{table}
Table \ref{table:va_ablation} and Table \ref{zeroshot_ablation} show that training without similarity predictor and random multimodal matching results in inferior performance compared to the baseline, demonstrating the effectiveness of the proposed method.
\section{Conclusion}
We have introduced MMVA, a novel framework for underexamined interactions among images, music, and musical captions using our IMEMNet-C dataset. MMVA addresses an unconventional multimodal training scenario: music-text pairs are one-to-one matched, while image-music and image-text pairs are aligned using continuous valence-arousal (VA) matching scores. By leveraging random multimodal matching based on VA scores, MMVA achieves state-of-the-art performance in VA prediction and demonstrates promising zero-shot capabilities in various zeroshot tasks. This makes MMVA particularly valuable for scenarios where obtaining one-to-one multimodal data is challenging, especially in music, where emotional content and subjective evaluation add complexity.
\paragraph{Limitations} While MMVA aligns modalities through emotional similarity, it may not fully capture other aspects of multimodal relationships, such as semantic alignment. Additionally, the inherent subjectivity of emotional perception could limit the model's generalizability across diverse cultural and demographic contexts. 

\section*{Acknowledgments}
This work is supported by the National Research Foundation of Korea (NRF) grant funded by the Korea government (MSIT)
 [RS-2024-00421203, RS-2024-00406127, 2021R1A2C3010887], Culture,  Sports  and  Tourism  R\&D  Program  through  the  Korea Creative  Content  Agency  grant  funded  by  the  Ministry  of  Culture,  Sports  and  Tourism  in  2024 (Project Name: Development of K-POP artist-centered video editing solution: customized multimodal AI model and generative asset, Project Number:  RS-2024-00399433, Contribution Rate: 50\%), and Artificial intelligence industrial convergence cluster development project funded by the Ministry of Science and ICT (MSIT, Korea) \& Gwangju Metropolitan City. 
 \bibliography{aaai25}

\end{document}